\documentclass[aps,prd,manuscript,amsfonts]{revtex4}

\begin{document}

\title{On the universality of the Carter and McLenaghan formula}

\author{Ion I. Cot\u aescu \\
{\it West University of Timi\c soara,}\\{\it V.
Parvan Ave. 4, RO-300223 Timi\c soara}}

\begin{abstract}
It is shown that the formula of the isometry generators of the spinor
representation given by Carter and McLenaghan is universal in the sense that
this holds for any representation either in local frames or even in natural
ones. The point-dependent spin matrices in natural frames are introduced for
any tensor representation deriving the covariant form of the isometry
generators in these frames.

Pacs: 04.20.Cv, 04.62.+v, 11.30.-j
\end{abstract}

\maketitle


\section{Introduction}

In general relativity there are many symmetries but only the isometries
(associated to Killing vectors) are able to produce conserved quantities via
Noether theorem. The physical fields minimally coupled to gravity take over
this symmetry transforming according to appropriate representations of the
isometry group. In the case of the scalar vector or tensor fields these
representations are completely defined by the well-known rules of the general
coordinate transformations since the isometries are in fact particular
automorphisms. On the other hand, the theory of spinor fields is formulated in
orthogonal local frames where the basis-generators of the spinor representation
were discovered by Carter and McLenaghan \cite{CML}

Some time ago we proposed a theory of external symmetry in the context of the
gauge covariant theories based on local orthogonal frames. We introduced there
new transformations which combine isometries and gauge transformations such
that the tetrad fields should remain invariant \cite{ES}. In this way we
obtained the external symmetry group which is nothing else that the universal
covering group of the isometry one. Moreover, we derived the general form of
the basis-generators of the representations of this group shoving that these
are given by a similar formula to that of Carter and McLenaghan.

In the present letter we should like to show that this formula holds even in
natural frames. To this end we define the spin matrices of the vector and
tensor representations in natural frames which are point-dependent and give
rise to the spin terms of the covariant form of our basis-generators. The
conclusion is that the Carter and McLenaghan formula is universal.

We start in the second section with a short introduction  to the tetrad gauge
covariant theories in which we defined our theory of external symmetry briefly
presented in the next section. The spin matrices in natural frames are
introduced in section three where the basis-generators of the vector and tensor
representations of the external symmetry group are obtained.

\section{Tetrad gauge covariance}

Let us consider the curved spacetime $M$ and a local chart (natural frame) of
coordinates $x^{\mu}$ (labeled by natural indices, $\mu, \nu,...=0,1,2,3$).
Given a gauge, we denote by $e_{\hat\mu}$ the tetrad fields that define the
local frames and by $\hat e^{\hat\mu}$ those of the corresponding coframes.
These have the usual orthonormalization properties, $\hat
e^{\hat\mu}_{\alpha}\, e_{\hat\nu}^{\alpha}=\delta^{\hat\mu}_{\hat\nu}$, $ \hat
e^{\hat\mu}_{\alpha}\, e_{\hat\mu}^{\beta}=\delta^{\beta}_{\alpha}$,
$e_{\hat\mu}\cdot e_{\hat\nu}=\eta_{\hat\mu \hat\nu}$, $\hat e^{\hat\mu}\cdot
\hat e^{\hat\nu}=\eta^{\hat\mu \hat\nu}$. The metric tensor $g_{\mu
\nu}=\eta_{\hat\alpha\hat\beta}\hat e^{\hat\alpha}_{\mu}\hat
e^{\hat\beta}_{\nu}$ raises or lowers the natural indices while for the local
indices ($\hat\mu, \hat\nu,...=0,1,2,3$) we have to use the Minkowski metric
$\eta=$diag$(1,-1,-1,-1)$. This metric remains invariant under the
transformations of its  gauge group, $O(3,1)$. This has as subgroup the Lorentz
group, $L_{+}^{\uparrow}$, of the transformations $\Lambda[A(\omega)]$
corresponding to the transformations $A(\omega)\in SL(2,\Bbb C)$ through the
canonical homomorphism. In the standard covariant parametrization, with the
real parameters $\omega^{\hat\alpha \hat\beta}=-\omega^{\hat\beta\hat\alpha}$,
the $SL(2,\Bbb C)$ transformation read
\begin{equation}\label{Aomega}
A(\omega)= e^{-\frac{i}{2}\omega^{\hat\alpha\hat\beta}
S_{\hat\alpha\hat\beta}}
\end{equation}
where $S_{\hat\alpha \hat\beta}$ are the covariant basis-generators of the
$sl(2,\Bbb C)$ Lie algebra. In fact, these are the principal spin operators
which generate the spin terms of other operators in the quantum theory. For
small values of $\omega^{\hat\alpha\hat\beta}$ the matrix elements of the
transformations $\Lambda$ in the local basis can be expanded as
$\Lambda^{\hat\mu\,\cdot}_{\cdot\,\hat\nu}[A(\omega)]=
\delta^{\hat\mu}_{\hat\nu} +\omega^{\hat\mu\,\cdot}_{\cdot\,\hat\nu}+\cdots$.

Assuming now that $M$ is orientable and time-orientable we can consider
$G(\eta)=L^{\uparrow}_{+}$  as the gauge group of the Minkowski metric $\eta$
\cite{WALD}. Then, the  fields with spin in local frames transform according to
finite-dimensional  linear representations, $\rho$, of the group $SL(2,\Bbb
C)$. In general, a matter field $\psi_{\rho}:M\to {\cal V}_{\rho}$ is defined
over $M$ with values in the vector space ${\cal V}_{\rho}$ of the
representation $\rho$ which is generally reducible. The covariant derivatives
of the field $\psi_{\rho}$,
\begin{equation}\label{der}
D_{\hat\alpha}^{\rho}= e_{\hat\alpha}^{\mu}D_{\mu}^{\rho}=
\hat\partial_{\hat\alpha}+\frac{i}{2}\, \rho(S^{\hat\beta\, \cdot} _{\cdot\,
\hat\gamma})\,\hat\Gamma^{\hat\gamma}_{\hat\alpha \hat\beta}\,,
\end{equation}
depend on the  connection coefficients in local frames,
$\hat\Gamma^{\hat\sigma}_{\hat\mu \hat\nu}=e_{\hat\mu}^{\alpha}
e_{\hat\nu}^{\beta}(\hat e_{\gamma}^{\hat\sigma} \Gamma^{\gamma}_{\alpha \beta}
-\hat e^{\hat\sigma}_{\beta, \alpha})$, which assure the covariance of the
whole theory under tetrad gauge transformations. This is the general framework
of the theories involving fields with half integer spin which the natural
frames are not dealing with.

\section{External symmetries in local frames}

A special difficulty in local frames is that the theory is no longer covariant
under isometries since these can change the tetrad fields that carry natural
indices. For this reason we proposed a theory of external symmetry in which
each isometry transformation is coupled to a gauge one able to correct the
position of the local frames such that the whole transformation should preserve
not only the metric but the tetrad gauge too \cite{ES}. Thus, for any isometry
transformation $x\to x'=\phi_{\xi}(x)=x+\xi^a k_a+...$, depending on the
parameters $\xi^a$ ($a,b,...=1,2...N$) of the isometry group $I(M)$, one must
perform the gauge transformation $A_{\xi}$ defined as
\begin{equation}\label{Axx}
\Lambda^{\hat\alpha\,\cdot}_{\cdot\,\hat\beta}[A_{\xi}(x)]=
\hat e_{\mu}^{\hat\alpha}[\phi_{\xi}(x)]\frac{\partial \phi^{\mu}_{\xi}(x)}
{\partial x^{\nu}}\,e^{\nu}_{\hat\beta}(x)
\end{equation}
with the supplementary condition $A_{\xi=0}(x)=1\in SL(2,\Bbb C)$.  Then the
transformation laws of our fields are
\begin{equation}\label{es}
(A_{\xi},\phi_{\xi}):\qquad
\begin{array}{rlrcl}
e(x)&\to&e'(x')&=&e[\phi_{\xi}(x)]\\
\hat e(x)&\to&\hat e'(x')&=&\hat e[\phi_{\xi}(x)]\\
\psi_{\rho}(x)&\to&\psi_{\rho}'(x')&=&\rho[A_{\xi}(x)]\psi_{\rho}(x)\,.
\end{array}
\qquad
\end{equation}
We have shown that the pairs $(A_{\xi},\phi_{\xi})$ constitute a well-defined
Lie group we called the external symmetry group, $S(M)$, pointing out that this
is just the universal covering group of $I(M)$ \cite{ES}. For small values of
$\xi^{a}$, the $SL(2,\Bbb C)$ parameters of $A_{\xi}(x)\equiv
A[\omega_{\xi}(x)]$ can be expanded as $\omega^{\hat\alpha\hat\beta}_{\xi}(x)=
\xi^{a}\Omega^{\hat\alpha\hat\beta}_{a}(x)+\cdots$, in terms of the functions
\begin{equation}\label{Om}
\Omega^{\hat\alpha\hat\beta}_{a}\equiv {\frac{\partial
\omega^{\hat\alpha\hat\beta}_{\xi}} {\partial\xi^a}}_{|\xi=0} =\left( \hat
e^{\hat\alpha}_{\mu}\,k_{a,\nu}^{\mu} +\hat e^{\hat\alpha}_{\nu,\mu}
k_{a}^{\mu}\right)e^{\nu}_{\hat\lambda}\eta^{\hat\lambda\hat\beta}
\end{equation}
which depend on the Killing vectors $k_a=\partial_{\xi_a}\phi_{\xi}|_{\xi=0}$
associated to $\xi^a$.

The last of Eqs. (\ref{es})  defines the operator-valued representation
$(A_{\xi},\phi_{\xi})\to T_{\xi}^{\rho}$ of the group $S(M)$ whose
transformations,
$(T_{\xi}^{\rho}\psi_{\rho})[\phi_{\xi}(x)]=\rho[A_{\xi}(x)]\psi_{\rho}(x)$,
leave the field equation invariant. Their basis-generators \cite{ES},
\begin{equation}\label{Sx}
X_{a}^{\rho}=i{\partial_{\xi^a} T_{\xi}^{\rho}}_{|\xi=0}=-i
k_a^{\mu}\partial_{\mu} +\frac{1}{2}\,\Omega^{\hat\alpha\hat\beta}_{a}
\rho(S_{\hat\alpha\hat\beta})\,,
\end{equation}
commute with the operator of the field equation and satisfy the commutation
rules $[X_{a}^{\rho}, X_{b}^{\rho}]=ic_{abc}X_{c}^{\rho}$ determined by the
structure constants, $c_{abc}$, of the $s(M)\sim i(M)$ algebras.  Eq.
(\ref{Sx}) can be put in covariant the form
\begin{equation}\label{CM}
X^{\rho}_{a}=-ik^{\mu}_{a}D_{\mu}^{\rho}+\frac{1}{2}\, k_{a\,
\mu;\,\nu}\,e^{\mu}_{\hat\alpha}\,e^{\nu}_{\hat\beta}\,
\rho(S^{\hat\alpha\hat\beta})\,,
\end{equation}
which represents a generalization to any representation $\rho$ of the famous
formula given by Carter and McLenaghan  for the spinor representation
$\rho_{spin}=(\frac{1}{2},0)\oplus (0,\frac{1}{2})$ \cite{CML}.

\section{Isometries in natural frames}

Whenever the spinors are absent, the matter fields are vectors or tensors of
different ranks  and the whole theory is independent on the tetrad fields
dealing with the natural frames only. Let us see what happens with our theory
of external symmetry in natural frames focussing on tensor fields. Any tensor
field, $\Theta$, transforms under isometries as $\Theta\to \Theta'
=T_{\xi}\Theta$, according to a  tensor representation of the group $S(M)$
defined by the well-known rule in natural frames
\begin{equation}
\left[\frac{\partial \phi^{\alpha}_{\xi}(x)}{\partial x_{\mu}}\frac{\partial
\phi^{\beta}_{\xi}(x)}{\partial x_{\nu}}...
\right]\left(T_{\xi}\Theta\right)_{\alpha\beta...}[\phi(x)]=\Theta_{\mu\nu...}(x)\,.
\end{equation}
Hereby one derives the basis-generators of the tensor representation,
$X_{a}=i\,\partial_{\xi_a}T_{\xi}|_{\xi=0}$, whose action reads
\begin{equation}\label{XT}
(X_a\,
\Theta)_{\alpha\beta...}=-i({k_a}^{\nu}\Theta_{\alpha\beta...;\,\nu}
+{k_a}^{\nu}_{~;\,\alpha}\Theta_{\nu\beta...}
+{k_a}^{\nu}_{~;\,\beta}\Theta_{\alpha\nu ...}...)\,.
\end{equation}
Our purpose is to show that the operators $X_a$  can be written in a form which
is equivalent to  Eq. (\ref{CM}) of Carter and McLanaghan.

We start with the vector representation $\rho_v=(\frac{1}{2},\frac{1}{2})$ of
the $SL(2, \Bbb C)$ group, generated by the spin matrices
$\rho_v(S^{\hat\alpha\hat\beta})$ which have the well-known matrix elements
\begin{equation}
[\rho_v(S^{\hat\alpha\hat\beta})]^{\hat\mu\,\cdot}_{\cdot\,\hat\nu}
=i(\eta^{\hat\alpha\hat\mu}\delta^{\hat\beta}_{\hat\nu}-
\eta^{\hat\beta\hat\mu}\delta^{\hat\alpha}_{\hat\nu})\,,
\end{equation}
in local bases. Furthermore, we define the point-dependent spin matrices in
natural frames whose matrix elements in the natural basis read
\begin{equation}\label{spin1}
(\tilde S^{\mu\nu})^{\sigma\,\cdot}_{\cdot\,\tau}
=e^{\mu}_{\hat\alpha}e^{\nu}_{\hat\beta}e^{\sigma}_{\hat\gamma} [\rho_v(
S^{\hat\alpha\hat\beta})]^{\hat\gamma\,\cdot}_{\cdot\,\hat\delta} \hat
e^{\hat\delta}_{\tau}=
i(g^{\mu\sigma}\delta^{\nu}_{\tau}-g^{\nu\sigma}\delta^{\mu}_{\tau})\,.
\end{equation}
In what follows we assume that these matrices represent the spin operators of
the vector representation in natural frames. We observe that these are the
basis-generators of the groups $G[g(x)]\sim G(\eta)$ which leave the metric
tensor $g(x)$ invariant in each point $x$. Since the representations  of these
groups  are point-wise equivalent with those of $G(\eta)$, one can show that in
each point $x$ the basis-generators $\tilde S^{\mu\nu}(x)$ satisfy the standard
commutation rules of the vector representation $\rho_v$ (but with $g(x)$
instead of $\eta$).

In general, the spin matrices of a tensor $\Theta$ of any rank, $n$, are the
basis-generators of the representation
$\rho_n=\rho_v^1\otimes\rho_v^2\otimes\rho_v^3\otimes....\otimes\rho_v^n$ which
read $\rho_n(\tilde S)=\tilde S^1\otimes I^2\otimes I^3... +I^1\otimes \tilde
S^2\otimes I^3...+....$. Using these spin matrices a straightforward
calculation shows that  Eq. (\ref{CM}) can be rewritten in natural frames as
\begin{equation}\label{Xvt}
X^{n}_{a}=-ik^{\mu}_{a}\nabla_{\mu}+\frac{1}{2}\, k_{a\, \mu;\,\nu}\,
\rho_n(\tilde S^{\mu\nu})\,,
\end{equation}
where $\nabla_{\mu}$ are the usual covariant derivatives. It is not difficult
to verify that the action of these operators is just that given by Eq.
(\ref{XT}) which means that $X_a^{n}$ are the basis-generators of a tensor
representation of rank $n$ of the group $S(M)$. Thus, it is clear that Eq.
(\ref{Xvt}) represents the generalization to natural frames of the Carter and
McLenaghan formula (\ref{CM}).

\section{Concluding remarks}

The principal conclusion here is that the Carter and McLenaghan formula is
universal since it holds not only in local frames but in natural frames too. In
local frames this formula can be written in the covariant form (\ref{CM}) or as
in Eq. (\ref{Sx}) where the orbital part (i.e. the first term) is completely
separated from the spin term (the second one). It is worth pointing out that
Eq. (\ref{Sx}) can not be rewritten in natural frames even though it might be
useful for analyzing the operators structure in the relativistic quantum
mechanics. This is because the connection in local frames can not be expressed
in terms of spin matrices $\tilde S$. Therefore, in natural frames we are
forced to use only the genuine Carter and McLenaghan formula (\ref{Xvt}).

This formula exhibits explicitly the point-dependent spin matrices of the
vector or tensor representations which may provide us with information about
the spin content of the various fields involved in theory. However, some
results could be surprising. Thus, for example, if a second order rank tensor
has the form $G_{\mu\nu}=g_{\mu\nu}F$ where $F$ is a scalar field then we
calculate $\rho_2(\tilde S^{\alpha\beta})G=0$ which indicates that the tensor
$G$ is spinless despite of the fact that it is of second rank. Consequently,
the metric tensor is also spinless as it seems to be natural since this is
covariantly  constant and invariant under isometries (such that $X^2_a\, g=0$).
Another example is presented in Ref. \cite{CCM} where we define the photon
polarization in de Sitter manifolds without to consider local frames.

Finally we note that, despite of our partial results, the problem of the spin
in general relativity remains open as long as we do not have an effective
theory similar to the Wigner theory of the induced representations of the
Poincar\' e group.

\subsection*{Acknowledgments}

We are grateful to Cosmin Crucean and Nistor Nicolaevici for interesting and
useful discussions on closely related subjects.

\end{document}